\documentclass[onecolumn,draftclsnofoot]{IEEEtran}
\usepackage{graphicx}
\usepackage{amsmath}
\usepackage{color}
\usepackage{cite}
\usepackage{graphicx}
\usepackage{enumerate}
\usepackage[caption=false]{subfig}
\usepackage{pbox}
\usepackage{array}
\usepackage{stfloats}

\begin{document}
\title{Design of a High-Performance High-Pass Generalized Integrator Based Single-Phase PLL}
\author{\IEEEauthorblockN{Abhijit Kulkarni and Vinod John}
                          \thanks{The authors are with the Department of Electrical Engineering, Indian Institute of Science, Bangalore - 560012.
                          They can be contacted through email at: abhiekulkarni@gmail.com and vjohn@ee.iisc.ernet.in. 
                          
                          A Kulkarni is presently affiliated with the Department of Electrical and Computer Engineering, University of Illinois at Chicago, Chicago IL 60607, USA.}}

\maketitle

\begin{abstract}
Grid-interactive power converters are normally synchronized to the grid using  phase-locked loops (PLLs). The performance of the PLLs is affected by the non-ideal conditions in the sensed grid voltage such as harmonics, frequency deviations and dc offsets in single-phase systems. In this paper, a single-phase PLL is presented to mitigate the effects of these non-idealities.  This PLL is based on the popular second order generalized integrator (SOGI) structure. The SOGI structure is modified to eliminate of the effects of input  dc offsets.  
The resulting SOGI structure has a high-pass filtering property. Hence, this PLL is termed as high-pass generalized integrator based PLL (HGI-PLL).  
It has fixed parameters which reduces the implementation complexity and aids in the implementation in low-end digital controllers. The HGI-PLL is shown to have least resource utilization among the SOGI based PLLs with dc cancelling capability. Systematic design methods are evolved leading to the design that  limits the unit vector THD to within $1\%$ for given non-ideal input conditions in terms of frequency deviation and harmonic distortion.
The proposed designs achieve the fastest transient response. The performance of this PLL has been verified experimentally. The results are found to agree with the theoretical prediction.

\end{abstract}


\section*{Nomenclature}
\addcontentsline{toc}{section}{Nomenclature}
\begin{IEEEdescription}[\IEEEusemathlabelsep\IEEEsetlabelwidth{$V_1,V_2$}]
\item[PLL] Phase-locked loop.
\item[SOGI] Second-order generalized integrator.
\item[SRF] Synchronous reference frame.
\item[HGI] High-pass generalized integrator.
\item[$k$] Gain in the HGI transfer functions.
\item[$\omega _0$] Nominal grid frequency ($2\pi50~rad/s$).
\item[$v_g$] Sensed grid voltage.
\item[$v_\alpha$,~$v_\beta$] In-phase and quadrature-phase outputs of HGI with $v_g$ as the input.
\item[$v_d$,~$v_q$] Rotating reference frame voltages corresponding to SRF transformation of $v_\alpha$,~$v_\beta$.
\item[$k_p$,~$k_i$] Proportional and integral gains of the PI controller.
\item[$\omega _e$] Estimated frequency of the HGI-PLL in rad/s.
\item[$\theta _e$] Estimated phase of the HGI-PLL in rad.
\item[$t_{s,srf}$] Settling time due to the embedded SRF-PLL in HGI-PLL.
\item[$t_{s,hgi}$] Settling time due to the HGI block.
\item[$t_{sd}$] Additive worst case settling time \\($t_{sd}= t_{s,srf}+ t_{s,hgi}$). 
\item[$\omega _{bw}$] Design bandwidth of the embedded SRF-PLL in rad/s.
\item[$f _{bw}$] Design bandwidth of the embedded SRF-PLL in Hz.
\item[$k_{opt,h}$] Optimum value of $k$ in HGI that gives the fastest settling time.
\item[$U$] Design limit on the unit vector THD in \%.
\item[$\Delta f$] Maximum frequency deviation considered in grid ($\Delta f = \pm 8\%$).
\item[$u_{thd}$] Unit vector THD in \%.
\item[$\mathbf{K_u}$] Set of $k$ satisfying $u_{thd} \leq 1\%$ for any given $f_{bw}$.
\end{IEEEdescription}

\section{Introduction}
Phase-locked loops (PLLs) are used in multiple applications: from miniature system on chip (SOC) to large grid-connected power converters. In SOCs, the PLLs are used for functions such as clock generation~\cite{pll_soc}. In grid-connected power converters such as  distributed generation (DG) systems and static compensators  (STATCOMs), PLLs are used for synchronization with the grid voltage~\cite{pll_dg1, pll_dg2, pv_dg_monitor, statcom2,statcom3,bsingh}. In this paper, the design and implementation aspects for PLL in single-phase grid connected power converters are discussed.

\begin{figure}[htbp]
\centering 
\includegraphics[scale=0.6]{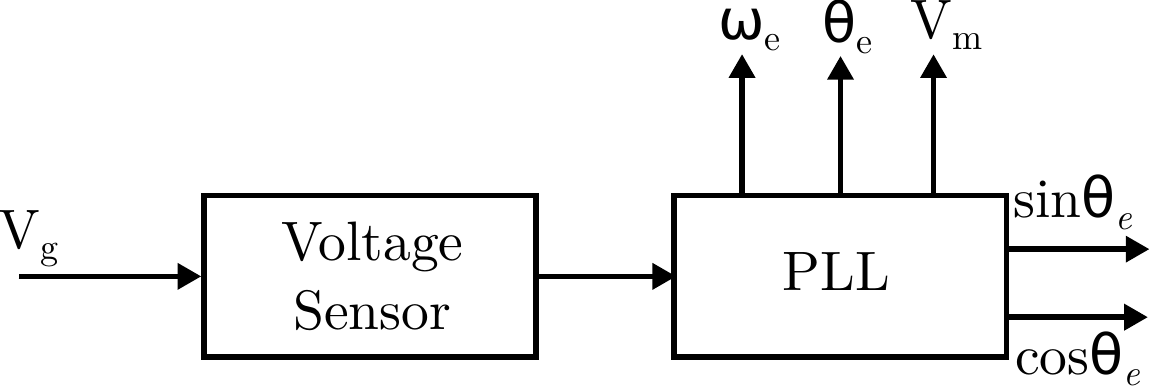} 
\caption{General structure of a PLL used in a grid-connected power converter. $V_g$ is the grid voltage. $\omega _e$, $\theta _e$ and $V_m$ are the frequency, phase and amplitude of grid voltage estimated by the PLL. $\sin \theta _e$ and $\cos \theta _e$ are the unit vectors.}
\label{pll_general} 
\end{figure}

The PLLs estimate the frequency, phase and amplitude of the grid voltage. They are used to generate unit amplitude sine and cosine signals synchronized with the grid voltage. These signals are called unit vectors~\cite{abhi1}. They are used for the reference signal generation in the closed-loop control of the power converters. The PLLs  are also used to monitor  the  disturbances in the grid voltage~\cite{pll_monitor1,karimi,pll_monitor3,pv_dg_monitor}.
Fig.~\ref{pll_general} shows the general schematic of a PLL used in the grid-synchronization  of a single-phase power converter.


The performance of the single-phase PLLs is affected by the non-ideal conditions in the grid voltage. These are: frequency deviation, harmonic distortion and dc offsets. In case of three-phase systems, there will be an additional non-ideality of unbalance in the three-phase voltages. It must be noted that under ideal conditions, the grid voltage has fixed frequency of either 50/60Hz, no harmonic distortion, no dc offsets and no unbalance in three-phase case. 

\subsection{Literature Survey of Existing PLL Structures}
Among the various PLLs proposed in literature, synchronous reference frame PLL (SRF-PLL) is a popular  PLL~\cite{blasko, abhi1} used in three-phase systems. It is very simple from design and implementation point of view. 
SRF-PLL forms the building block of many  PLLs  for both three-phase and single-phase applications~\cite{srfpll_based1,srfpll_based3,ghoshal2015,ciobotaru1,ciobotaru2,abhi_iecon,adaptive_sogi_blaabjerg}. Second-order-generalized-integrator (SOGI) based single-phase PLLs~\cite{ciobotaru1,ciobotaru2,abhi_iecon,adaptive_sogi_blaabjerg} are low-complexity single-phase PLLs that use an embedded SRF-PLL. The basic SOGI-PLL was first introduced in~\cite{ciobotaru1}. This PLL will have sinusoidal ripple errors in the estimated frequency when the input contains harmonics and dc offsets. These ripple errors can also occur when the input frequency changes from the nominal value~\cite{ciobotaru1}. Adaptation of the SOGI parameters is suggested to overcome the problem due to input frequency deviation~\cite{ciobotaru1,ciobotaru2,adaptive_sogi_blaabjerg,adaptive_sogi1}. The adaptive SOGI-PLLs  have higher design and implementation complexity. 

Input to the single-phase SOGI based PLLs can have a dc offset due to factors such as sensor dc offsets, dc offsets from the analog-to-digital controllers (ADC) and mismatch in the semiconductor device switching in practical power converters. As the basic SOGI structure cannot eliminate dc offsets, the embedded SRF-PLL will have a dc offset in its input. This can result in a   serious problem of dc injection to the grid~\cite{abhi_iet}. DC injection to the grid is undesired~\cite{dc_offset_effect_pow_del}  and it is to be limited to be less than $0.5\%$ of the rated current of the power converter as per the grid interconnection standard IEEE 1547-2003~\cite{ieee1547}.

The problem of dc offsets in the basic SOGI-PLL is mitigated in~\cite{ciobotaru2}, where the design method  is based on heuristic approach. Also, the effect of input harmonics is not  quantified in the work. Multiple cascaded SOGI based frequency locked loop (FLL) is proposed in~\cite{multiple_cascade}. This is proposed for  three-phase systems. Design optimizations considering the response time have not been analyzed. 
Cascading of the SOGI blocks increases the implementation complexity in terms of increased computation time or digital resource utilization. In~\cite{robust_shinnaka}, a modified high-pass based SOGI structure is studied for   a robust adaptive PLL. This PLL contains additional non-linear functions compared to other SOGI based PLLs such as~\cite{ciobotaru1,ciobotaru2,abhi_iecon,adaptive_sogi_blaabjerg} and the response to transients reported in the paper is  slow in the order of tens of fundamental cycles. The work in~\cite{pow_del_dc_off1} estimates the grid frequency and amplitude correctly when the input contains dc offsets. However, the phase estimation is affected by the input dc offsets.


\subsection{Present Work}
In this paper, a modified SOGI-PLL is presented. The modified SOGI has full dc offset rejection capability and includes a high-pass based filter structure. Hence, this PLL is termed as high-pass generalized integrator based PLL (HGI-PLL). The outputs of the HGI are given as inputs to the embedded SRF-PLL block. 

\begin{figure*}[t]
\centering 
\includegraphics[scale=0.66]{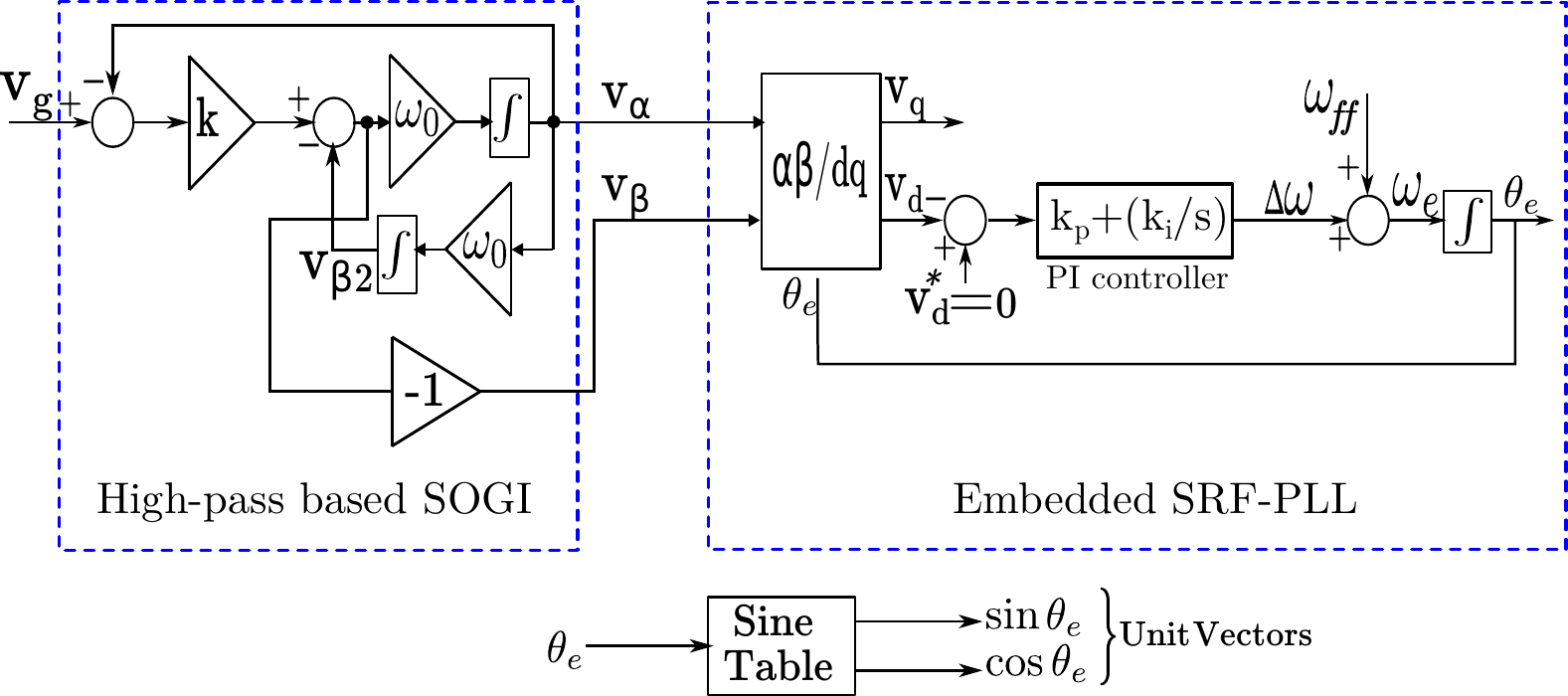} 
\caption{Structure of  high-pass generalized integrator based PLL (HGI-PLL).}
\label{hgipll_str} 
\end{figure*}

The structure of the HGI-PLL is shown in Fig.~\ref{hgipll_str}. 
This is a fixed parameter or non-adaptive PLL which helps to keep its implementation simple. Its performance is affected by the frequency deviations  as well as harmonics in the input voltage. These non-ideal conditions of frequency deviations and harmonics result in unit vector harmonic distortion~\cite{abhi_iecon}. This is undesirable as the unit vectors are used for reference generation and are expected to have minimal harmonic distortion.  
Hence, the HGI-PLL must be designed such that the the unit vector distortion is minimal for the non-ideal grid conditions of frequency deviation and harmonic distortion. Another desirable performance parameter is the fast settling time. The HGI-PLL design must consider these factors of minimal unit vector distortion and  fast response time for given worst case non-idealities in the grid voltage.

Novel systematic designs  are evolved for the HGI-PLL in this paper. For a given worst case frequency deviation in the input, a design approach is proposed which results in the fastest response for a given constraint on the unit vector THD. For example, for a worst case setting of $\pm8\%$ frequency deviation, the HGI-PLL is designed to have fastest response while limiting the unit vector THD to be less than $1\%$. This design is evolved into a design procedure considering two constraints, namely, the worst case frequency deviations {and} harmonic distortions in the input. This design achieves fastest response for HGI-PLL while limiting the unit vector THD to be within $1\%$ for the given worst-case input conditions.


The HGI-PLL with the proposed designs achieves very good transient and steady-state performance. The practical settling time is shown to be less than $30ms$.  This PLL has the least resource utilization in the digital implementation among the SOGI based PLLs with dc cancelling capability. The performance of this PLL has been compared with analysis and simulation and is  validated using experimental results for various steady-state and transient operating conditions.


\section{Structure and Design Considerations of HGI-PLL }
\subsection{Structure}
\begin{figure}[htbp]
\centering
\includegraphics[scale=0.38]{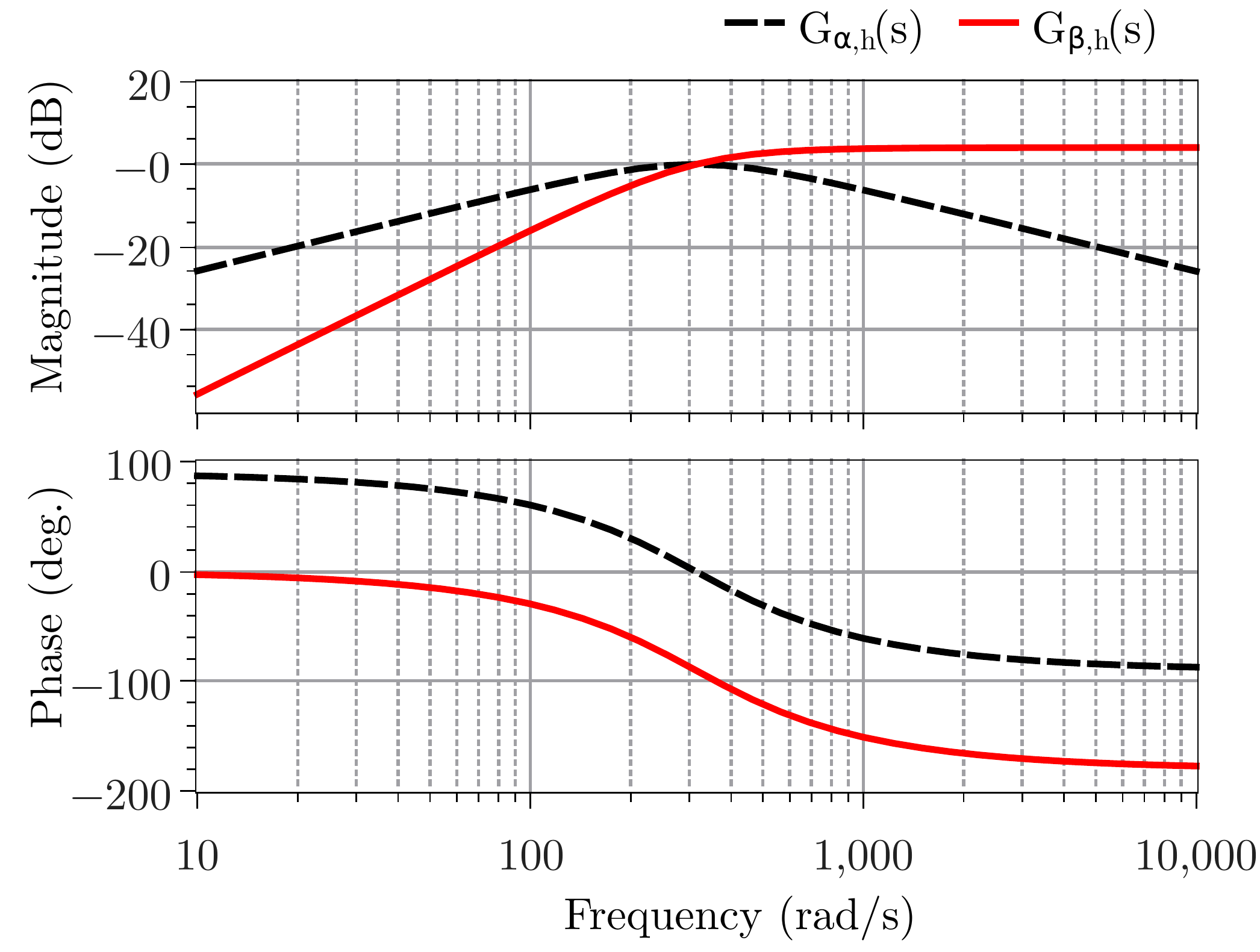}
\caption{Bode plot of the transfer functions  of HGI-PLL when $k=1.6$. }
\label{hgipll_bode}
\end{figure}

The HGI-PLL produces two quadrature signals $v_\alpha$ and $v_\beta$ from the input sensed grid voltage $v_g$. The HGI is a modified SOGI filter structure which generates $v_\alpha$ and $v_\beta$. The transfer functions realized by the HGI are as follows.
\begin{align}
G_{\alpha,h}(s)&=\frac{v_\alpha}{v_g}=\frac{ks\omega_0}{s^2+k\omega_0s+\omega_0^2}
\label{gatf_h} \\
G_{\beta,h}(s)&=\frac{v_\beta}{v_g}=-\frac{ks^2}{s^2+k\omega_0s+\omega_0^2}\label{gbtf_h}
\end{align}
It can be verified that both the transfer functions in (\ref{gatf_h}) and (\ref{gbtf_h}) have zero gain at dc. When the input voltage is sinusoidal with frequency $\omega_0$, it can be seen that the $v_\alpha$ and $v_\beta$ are balanced quadrature signals. This is evident from the bode plots of the two transfer functions in Fig.~\ref{hgipll_bode}.

As it can be seen from Fig.~\ref{hgipll_bode}, the transfer function in (\ref{gbtf_h}) is a high-pass filter. In a basic SOGI-PLL~\cite{ciobotaru1}, the transfer function to generate $v_\beta$ is a low-pass filter. Hence, it does not block any dc offset in the input voltage. As a result, the embedded SRF-PLL will have dc offsets in its input which in turn  lead to dc offsets in the PLL unit vectors~\cite{abhi_iet}. In HGI-PLL, this is mitigated by making use of a dc blocking high-pass filter as shown in Fig.~\ref{hgipll_str}.

\subsection{Design Considerations}
The HGI-PLL has three design parameters: the gain $k$ in the HGI transfer functions, $k_p$ and $k_i$ of the PI controller transfer function as shown in Fig.~\ref{hgipll_str}. In adaptive SOGI based PLLs, the term $\omega_0$ used in the transfer functions is replaced by the estimated frequency $\omega_e$ of the PLL~\cite{adaptive_sogi1}. In the present implementation, $\omega_0$ is constant and is equal to the nominal grid frequency in rad/s. By keeping $\omega_0$ fixed, the implementation is simplified. Use of fixed parameters helps in arriving at a systematic design method and hence the response time can be optimized. 

The main disadvantage of fixed parameter SOGI based PLLs is the fact that frequency deviations in the input voltage will cause unequal amplitudes in the $v_\alpha$ and $v_\beta$. This is clear from Fig.~\ref{hgipll_bode} also. The unequal amplitudes result in the application of a negative sequence component to the embedded SRF-PLL~\cite{abhi_iecon}. This results in double harmonic ripple in the estimated frequency~\cite{blasko} and hence, harmonic distortion in the unit vectors~\cite{abhi1}. Harmonic distortion in the unit vectors is highly undesirable because the current references generated using them will also become distorted. As the current controllers used normally have low-pass filtering characteristics with high bandwidth~\cite{current_control_ref1}, the resulting grid current will be distorted. Grid interconnection standards such as IEEE 1547-2003~\cite{ieee1547} have defined limits on the harmonic injection to the grid. The PLL design should ensure that these limits are not exceeded. 
This can be achieved by carefully selecting the bandwidth of the embedded SRF-PLL. Lower the bandwidth, better will be the harmonic attenuation for the unit vectors when the input has frequency deviations~\cite{abhi_iecon}. The response time of the embedded SRF-PLL is inversely related to its bandwidth. The relation between settling time $t_{s,srf}$ and its bandwidth $\omega_{bw}$ in rad/s can be approximated as~\cite{abhi_iet},
\begin{equation}
t_{s,srf}=4/\omega_{bw} \label{tsrf}
\end{equation} 
There is a tradeoff between the response time and harmonic attenuation. Hence, it is important to arrive at a design bandwidth of the embedded SRF-PLL such that the response time is fastest for a given constraint on the unit vector harmonic distortion or THD. This approach of the PLL parameter design is followed for the basic SOGI-PLL in~\cite{abhi_iecon}.

When the grid voltage has transient amplitude or phase jumps, the filters used in the HGI will take a known amount of time to settle. The transients in the HGI output will affect the inputs $v_\alpha$ and $v_\beta$ to the embedded SRF-PLL. 
This results in temporary errors in estimation of frequency and phase till the HGI outputs settle to the correct values. Hence, the overall settling time is also dependent on the design parameter $k$ of the HGI block. It must be designed to have a fast response during transient changes in the grid voltage. Let the overall settling time due to the HGI transfer functions be defined as,
\begin{equation}
t_{s,hgi}=\max(t_{s\alpha}(k),t_{s\beta}(k)) \label{ts,hgi_def}
\end{equation}
In (\ref{ts,hgi_def}), $t_{s\alpha}(k)$ and $t_{s\beta}(k)$ are the settling times of the transfer functions in (\ref{gatf_h}) and (\ref{gbtf_h}) for step change. These settling time values depend on the parameter $k$. 

As the HGI-PLL has a cascade of the HGI block and embedded SRF-PLL block, the worst-case additive settling time is given by the sum of the settling times of the HGI blocks and the embedded SRF-PLL. This worst-case additive settling time is termed as $t_{sd}$ and it is given by,
\begin{equation}
t_{sd}=t_{s,hgi}+t_{s,srf} \label{tsd_def}
\end{equation}

The grid voltage at the point of common coupling (PCC) normally contains lower order harmonics due to the presence of non-linear loads in the system and finite grid impedance~\cite{pcc_dist1,pcc_dist2}. The transfer function in (\ref{gatf_h}) has bandpass configuration centred around nominal fundamental frequency $\omega_0$. Hence, there will be attenuation of harmonics by this transfer function. However, as it can be seen from Fig.~\ref{hgipll_bode}, the transfer function (\ref{gbtf_h}) has high-pass characteristic and hence it cannot attenuate any harmonics in the input. The embedded SRF-PLL has a low-pass characteristic and can attenuate the harmonics. Hence, to sufficiently attenuate the input harmonics, both $k$ and PI controller parameters of the embedded SRF-PLL will have to be selected carefully as proposed in Section III of the paper. This also has an effect on the overall settling time, which is to be minimized as explained in Section III. 

\section{Design of the HGI-PLL}
The design of HGI-PLL discussed in the following subsections III-A  will be referred to as as minimum $t_{sd}$ design (MTSD design). Note that $t_{sd}$ is the additive settling time defined in (\ref{tsd_def}) which is the sum of settling time of the HGI block and the settling time of embedded SRF-PLL block. This design selects the values of $k$ and bandwidth considering the expected range of frequency deviations alone. The objective of this design is to select the PLL parameters that lead to shortest additive settling time for a specified limit on unit vector THD considering  worst case frequency deviations in the input. This design is analyzed for its harmonic attenuation capability and the resulting limitations are then explained. The MTSD design forms the basis for  a complete design method  proposed in  Section III-B. This is called as harmonic constrained minimum $t_{sd}$ design (HC-MTSD design),  because the design parameters are selected to achieve a limit on the unit vector THD when the input contains both frequency deviation and harmonic distortion. In other words, this design optimizes the response time of HGI-PLL without exceeding the THD limit on the unit vectors when the input voltage contains both frequency deviations and harmonic distortion.  

\subsection{MTSD Design for HGI-PLL}

\subsubsection{Selection of the parameter $k$ for MTSD design}
\begin{figure}[t]
\centering 
\includegraphics[scale=0.23]{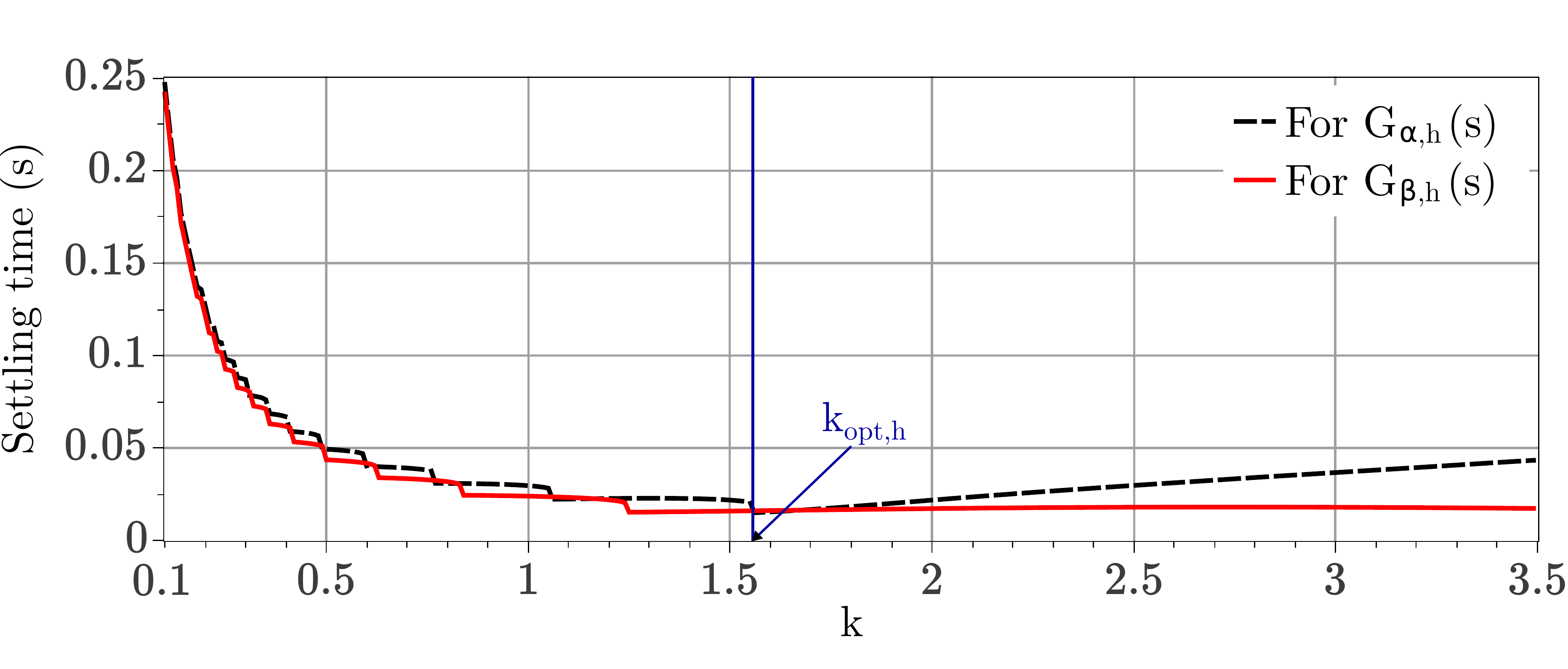} 
\caption{Variation of  settling time of the HGI transfer functions versus $k$  for MTSD design.}
\label{k_vs_tsett_pllh} 
\end{figure}

The parameter $k$ is selected such that the transfer functions of HGI-PLL give the fastest settling time.  
The step response settling time of the transfer functions (\ref{gatf_h}) and (\ref{gbtf_h})  as a function of $k$ is determined using simulation.
The variation of $k$ with 2\% step-response settling time~\cite{ogata} for HGI-PLL is shown  in Fig.~\ref{k_vs_tsett_pllh}. As it can be observed, there is a value of $k$ that gives the fastest response, corresponding to the minimum settling time of the SOGI block in Fig.~\ref{hgipll_str}.

From Fig.~\ref{k_vs_tsett_pllh},  the optimal value of $k$ is determined to be,
\begin{equation}
k_{opt,h}=1.56 \label{kopt}
\end{equation}
For this value of $k_{opt,h}$, the settling times for $v_\alpha$ and $v_\beta$ are $14.91ms$ and $15.97ms$ respectively for a $50Hz$ system. Hence, the combined worst-case settling time ($t_{s,hgi}$) would be the maximum of the two settling times, that is,
\begin{equation}
t_{s,hgi}=15.97ms \approx 16ms \label{thgi_max}
\end{equation}
Thus the worst case settling time of $16ms$  is less than one fundamental cycle.

Selection of $k=k_{opt,h}$ results in the fastest response of the HGI blocks for any changes in the input. The harmonic attenuation will be fixed based on the selected value of $k$. The remaining design parameter is the bandwidth of the embedded SRF-PLL.

\subsubsection{Selection of the bandwidth of embedded SRF-PLL for MTSD design}
As explained in Section II-B, the frequency deviations in the input result in unit vector distortion. To minimize the unit vector THD, the  approach is to adjust the bandwidth of the SRF-PLL to attenuate the resulting ripple in the $v_d$ and estimated frequency. Let the limit on the unit vector THD be $U=1\%$. Hence, for a given frequency deviation of $\Delta f$, the SRF-PLL bandwidth must be chosen to be such that  
the unit vector THD is $u_{THD}\leq U~(\%)$.  

\begin{figure}[htbp]
\centering 
\includegraphics[scale=0.46]{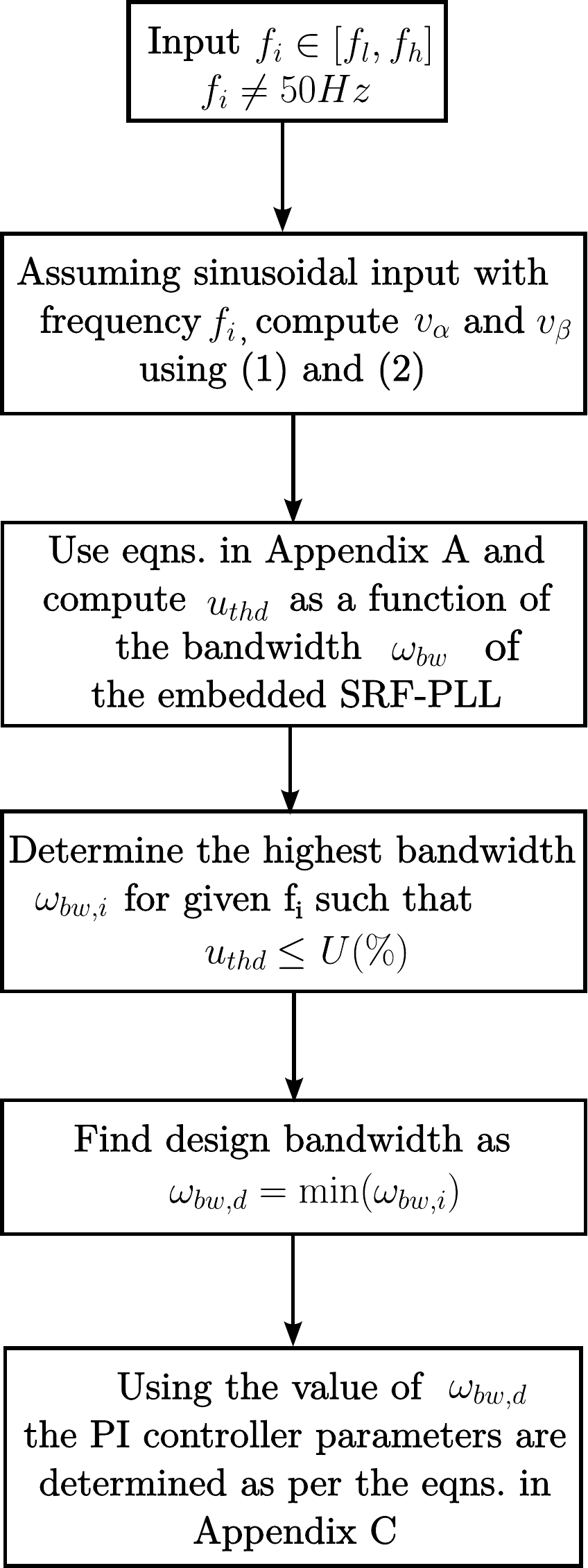} 
\caption{Flowchart to determine the design bandwidth of the embedded SRF-PLL for different levels of frequency deviations in the input.}
\label{flowchart_bw_uthd1} 
\end{figure}

The analytical expressions derived in~\cite{abhi1} are used to determine the unit vector THD as a function of the bandwidth of the embedded SRF-PLL for any frequency deviation in the input. The relevant expressions derived in~\cite{abhi1} are included in Appendix A for  quick reference. The procedure to determine the design bandwidth  is given in the flowchart in Fig.~\ref{flowchart_bw_uthd1}. The frequency range is swept from $f_i=[f_l,f_h]$.  For a frequency deviation of $\Delta f=\pm8\%$, the frequency is swept between $f_l=46Hz$ and $f_u=54Hz$ in a $50Hz$ system. Step size of  $2Hz$ has been used to sweep this range. 

\begin{figure}[htbp]
\centering 
\includegraphics[scale=0.23]{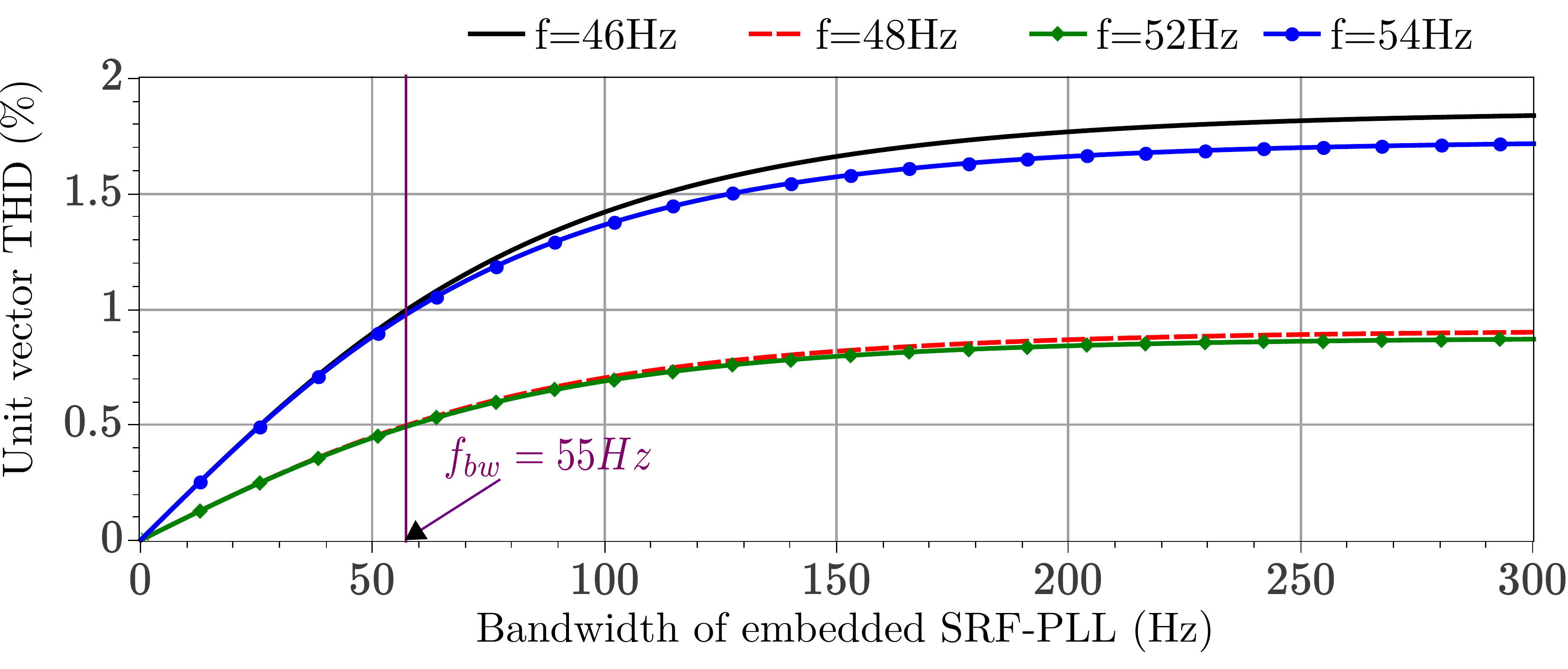} 
\caption{Variation of unit vector THD with bandwidth for various grid frequency values assuming sinusoidal grid voltage for HGI-PLL.}
\label{bw_uthd1} 
\end{figure}
This method is illustrated in Fig.~\ref{bw_uthd1}.
Consider when the input frequency is $46Hz$. The equations (\ref{aphi})--(\ref{mx_2}) in Appendix A are used and the unit vector THD versus bandwidth are plotted in Fig.~\ref{bw_uthd1}. The highest bandwidth is chosen for this case as per the flowchart to give a unit vector THD of at most $U=1\%$. This corresponds to the design bandwidth of $55Hz$. Similarly, the bandwidth is determined for the remaining three cases in Fig.~\ref{bw_uthd1}. It can be observed from the Fig.~\ref{bw_uthd1} that the bandwidth must be chosen to be $55Hz$ so that the unit vector THD is less than $1\%$ for the entire range of frequency deviations from $46Hz$ to $54Hz$. 

Thus, the  bandwidth $\omega_{bw,d}=2\pi55~rad/s$ will limit the unit vector THD to be less than $1\%$ even for a frequency deviation of upto $\pm8\%$ in the input voltage. Once the design bandwidth is known, the PI controller parameters of the embedded SRF-PLL can be determined using the design equations given in works such as~\cite{blasko, abhi_iet}. The final expressions (\ref{kpeqn}) and (\ref{kieqn}) used to compute these PI controller parameters are provided in Appendix C.  The settling time of the embedded SRF-PLL for this design bandwidth is given by,
\begin{equation}
t_{s,srf}=4/\omega_{bw,d}=11.6ms \label{design1_tsrf}
\end{equation}

\subsubsection{Summary of the MTSD design}
The $k$ in the HGI transfer functions is selected to achieve the fastest response. The bandwidth of the embedded SRF-PLL is the highest for given frequency deviation and constraint on unit vector THD. Hence for the given constraints, the embedded SRF-PLL also has the fastest response. The net worst-case additive settling time using (\ref{thgi_max}) and (\ref{design1_tsrf}) is given by,
\begin{equation}
t_{sd}=t_{s,hgi}+t_{s,srf}=16ms + 11.6ms = 27.6ms \label{add_settl_time}
\end{equation} 
Thus the proposed design results in a worst-case settling time of less than 1.5 fundamental cycles. The actual settling time is lesser than this value as the transients in HGI and embedded SRF-PLL occur simultaneously. This is shown in the Section V with experimental results.

\subsubsection{Effect of input harmonics on the MTSD design} The effect of the input harmonics is quantified analytically for this design as it does not include the input harmonic distortion  during the design process. A known amount of input THD is considered. The individual harmonics considered are the low order odd harmonics upto the ninth harmonic. The harmonic amplitude is considered to be inversely proportional to its harmonic order. That is,
\begin{equation}
\frac{v_{h,i}}{v_{h,j}}=\frac{j}{i} \label{harmonic_ratio}
\end{equation}
Hence, the fifth harmonic has an amplitude of $3/5$ times third harmonic, the seventh harmonic is $3/7$ times third harmonic and the ninth harmonic is $3/9$ times third harmonic. Based on this individual harmonics are calculated for a given THD. 

\begin{figure}[htbp]
\centering 
\includegraphics[scale=0.42]{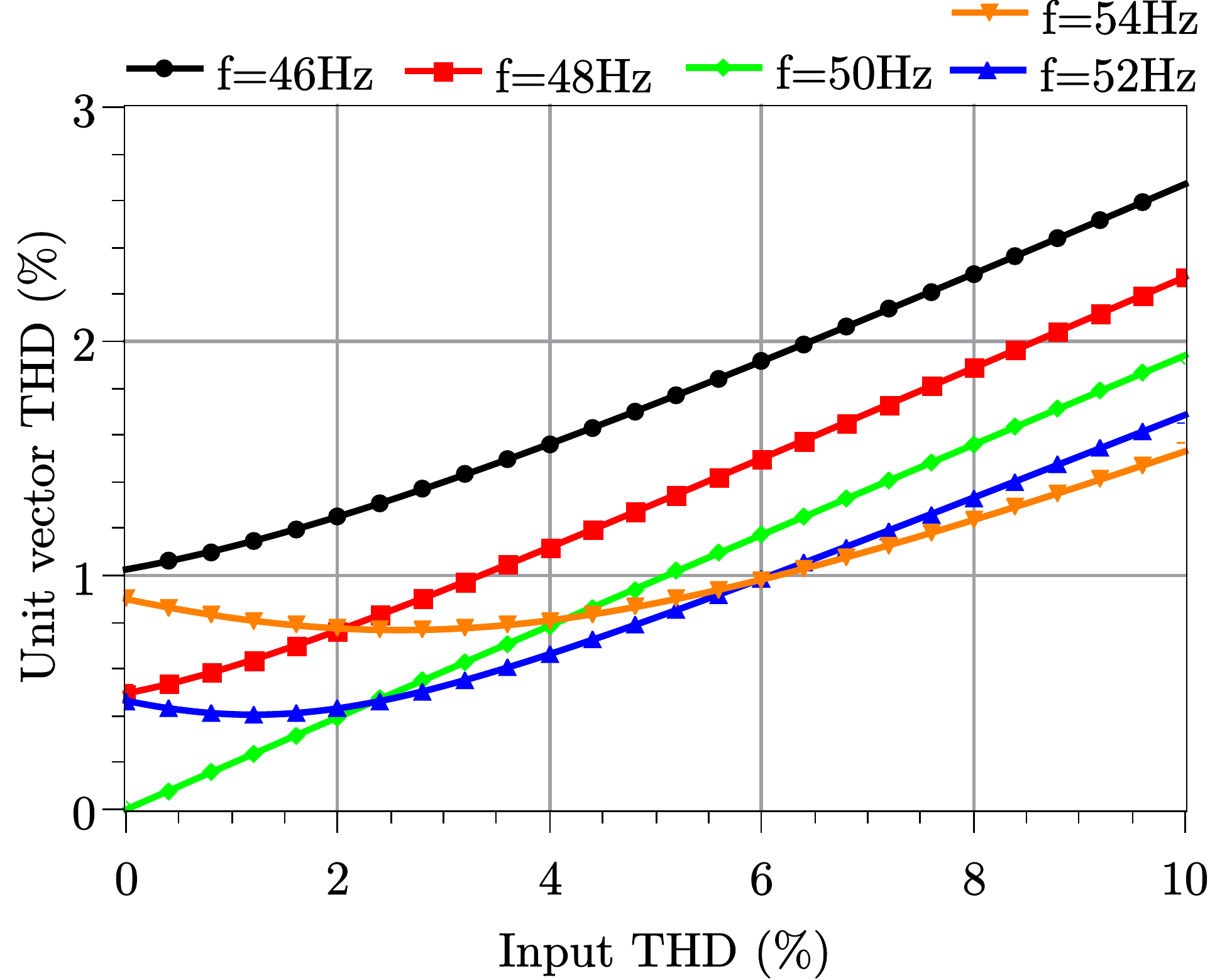}
\caption{Variation of unit vector THD with input THD for upto $\pm8\%$ frequency deviation in the input for MTSD design of HGI-PLL.}
\label{uthd_vs_ipthd1} 
\end{figure}

For each input THD, the resulting harmonics in the outputs of the transfer functions in (\ref{gatf_h}) and (\ref{gbtf_h}) are determined analytically. The resulting distorted $v_\alpha$ and $v_\beta$ are input to the embedded SRF-PLL. Hence, using the analytical expressions derived in~\cite{abhi1}, the unit vector THD is determined as a function of the input THD and frequency deviations. The unit vector THD can be evaluated from the phasor sum of individual unit vector harmonic distortion given by (\ref{aphikk})--(\ref{mx_h}) in Appendix B.  The resulting variation between the unit vector THD and the input THD for the MTSD design of the HGI-PLL is given in Fig.~\ref{uthd_vs_ipthd1}.

As it can be seen from Fig.~\ref{uthd_vs_ipthd1}, when the input has a THD of $5\%$, the unit vector THD is equal to $1.7\%$ when the input has frequency of $46Hz$. Thus, when the input THD is assumed to be a worst-case value of $5\%$~\cite{pcc_dist2,ieee1159}, the unit vector THD can exceed the $1\%$ limit.
To meet the objective of limiting the unit vector THD  when the input contains both frequency deviation and harmonics, an enhanced design procedure is evolved for the HGI-PLL. This is termed as the HC-MTSD design. This design limits the unit vector THD to be within  a limit of $U=1\%$ and achieves the fastest additive settling time while considering both frequency deviation and input THD. 
This design method is explained in the following subsection.

\subsection{HC-MTSD Design for HGI-PLL}
The objectives of this design can be defined as follows.
\begin{align}
\text{minimize}~t_{sd}=g(f_{bw},k) \label{detune_statement}
\end{align}
such that
\begin{align}
u_{thd}\leq 1\% \label{detune_uthd_constr}
\end{align}
given that the frequency deviation is $\Delta f (\%)$ and the input THD is $H(\%)$.

The additive settling time $t_{sd}$ is a function of the HGI-PLL parameters $f_{bw}$ and $k$. This function is labelled as $g(f_{bw},k)$ in (\ref{detune_statement}). Unit vector THD $u_{thd}$ is affected by both $f_{bw}$ and $k$. 
The upper limit on bandwidth for the solution of (\ref{detune_statement}) is considered to be  $55Hz$ based on the MTSD design. This is  because, from fundamental frequency deviation point of view, any higher $f_{bw}$ would result in unit vector THD of higher than 1\% as indicated in Fig.~\ref{bw_uthd1} for upto $\pm 8\%$ deviation in grid fundamental frequency.  The lower limit is considered to be $20Hz$. Theoretically, the lower limit on the bandwidth can be  close to zero. 

The optimum solution for (\ref{detune_statement}) considering the constraint (\ref{detune_uthd_constr}) and   the inputs of $\Delta f=\pm 8\%$ and $H=5\%$ is determined within the bandwidth range of $[f_{min},~ f_{max}]$. The range of $k$ considered is $[k_{min},~ k_{max}]$. These are listed as follows.
\begin{align}
&f_{min}=20Hz~~\text{and}~~f_{max}=55Hz \notag \\
&k_{min}=0.1~~\text{and}~~k_{max}=4 \label{detuned_input_range}
\end{align} 
The range of $k$ considers a wide possible design selection range as in Fig.~\ref{k_vs_tsett_pllh}. This range includes the $k_{opt,h}$ for the MTSD design specified in (\ref{kopt}).

\begin{figure}[htbp]
\centering 
\includegraphics[scale=0.43]{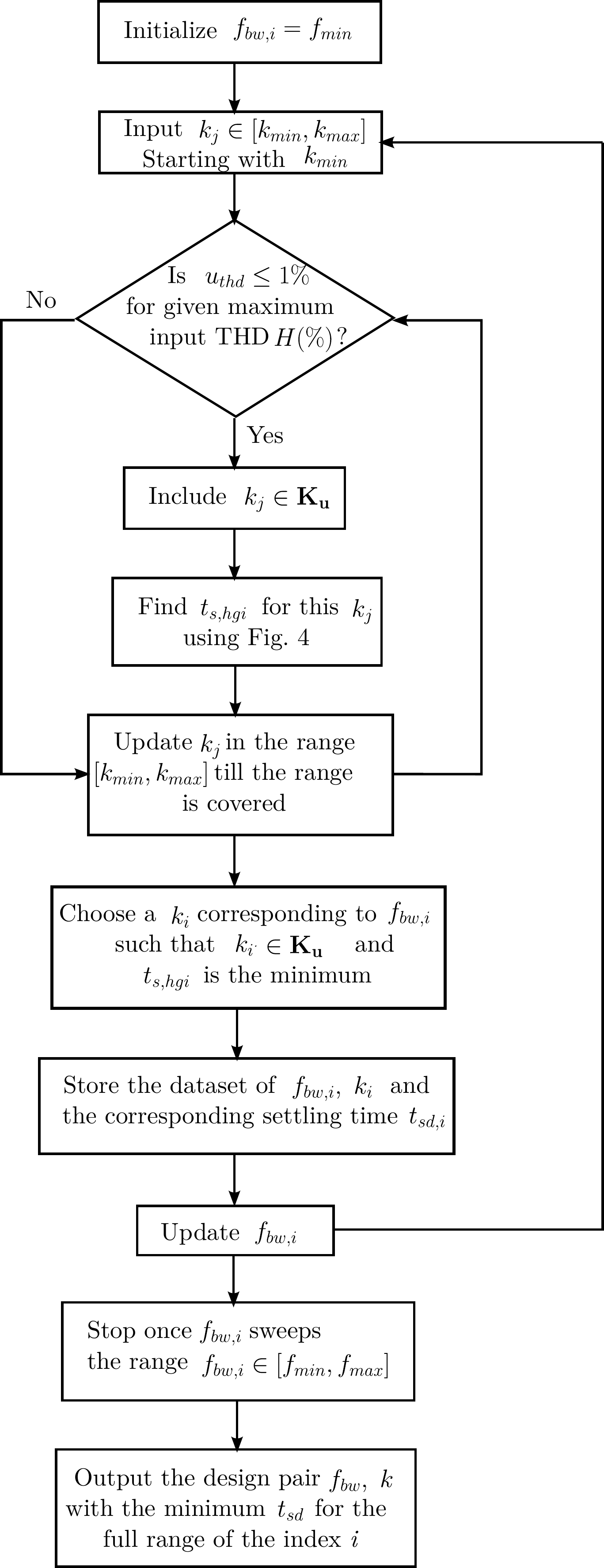} 
\caption{Flowchart showing the steps in HC-MTSD design of HGI-PLL.}
\label{hc_mtsd_flowchart} 
\end{figure}

The procedure for HC-MTSD design is explained using a flowchart in Fig.~\ref{hc_mtsd_flowchart}.
This is qualitatively explained as follows. For every bandwidth value in the range considered, the set of $k$ satisfying unit vector THD to be less than $1\%$ is determined. This set is designated as $\mathbf{K_u}$ in the flowchart in Fig.~\ref{hc_mtsd_flowchart}. In this set $\mathbf{K_u}$, the value of $k$ that gives fastest response time for the HGI is determined. This is done using a procedure similar to the plot   in Fig.~\ref{k_vs_tsett_pllh}. This value of $k$ is selected as the corresponding value to the bandwidth chosen. The additive settling time of of the HGI-PLL is computed for this pair of bandwidth and $k$. This is repeated for a large number of bandwidth values in the range specified in (\ref{detuned_input_range}). For each bandwidth, a corresponding $k$ and settling time are determined. The pair with least additive settling time is selected as the design value based on the objective in (\ref{detune_statement}).

\begin{figure}[htbp]
\centering
\includegraphics[scale=0.23]{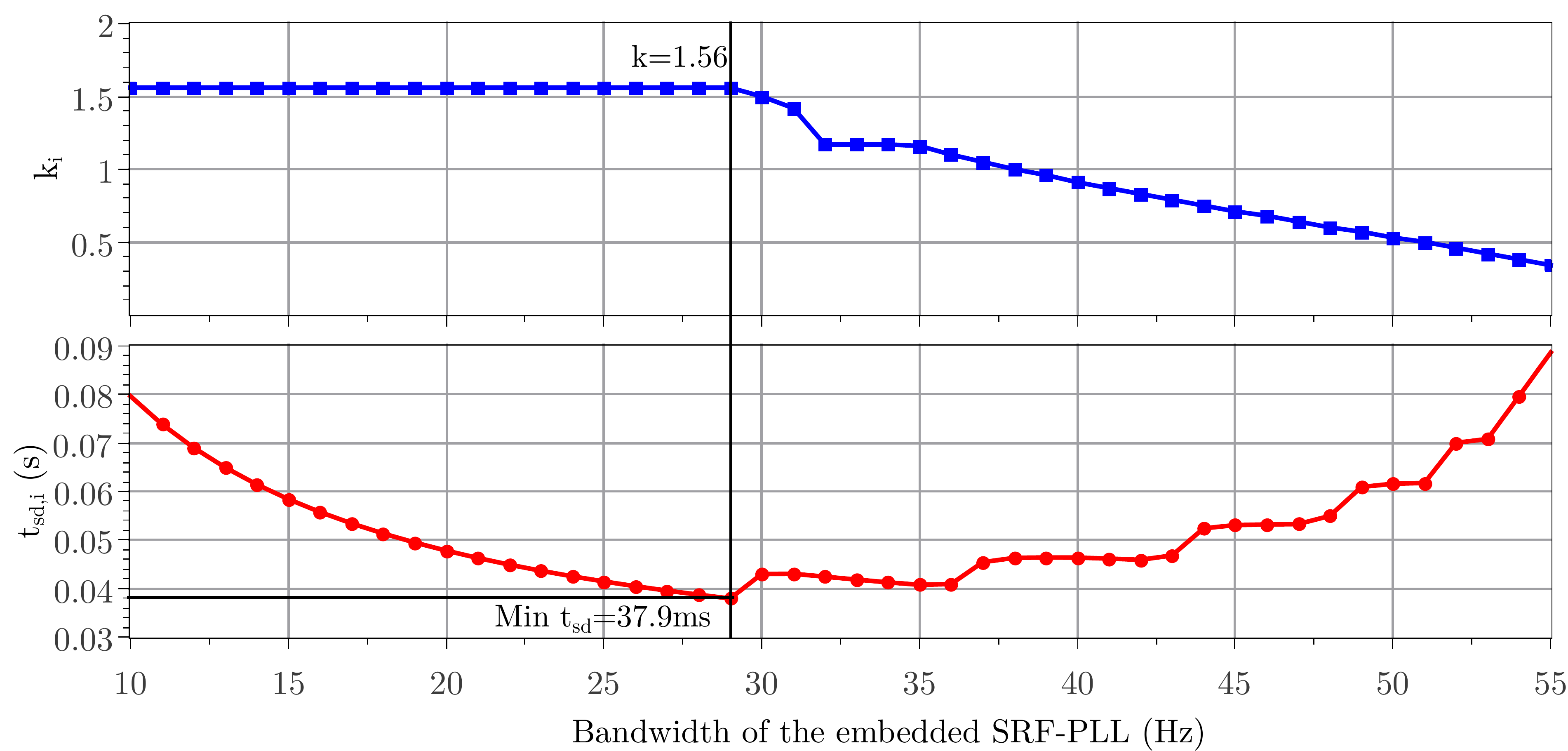} 
\caption{HC-MTSD design for HGI-PLL. Variation of $k_i$ with bandwidth (top trace), and variation of additive settling time $t_{sd,i}$ with bandwidth (bottom trace). }
\label{detune_design_pllh} 
\end{figure}

The solution to this method is illustrated graphically as follows.  The valid solution points contain three components which are $f_{bw,i}$, $k_i$ and corresponding $t_{sd,i}$. In the top trace in Fig.~\ref{detune_design_pllh}, $k_i$ is plotted versus $f_{bw,i}$. In the bottom trace, $t_{sd,i}$ is plotted versus $f_{bw,i}$. It can be observed that a minimum $t_{sd}$ is obtained for $f_{bw}=29Hz$ in the bottom trace. The corresponding $k=1.56$. This is the optimal pair of $f_{bw}$ and $k$ and gives a settling time of $37.9ms$ which is less than two fundamental cycles in a 50Hz system. The plot in Fig.~\ref{detune_design_pllh} is for a worst case input THD of $H=5\%$ and frequency deviation of $\Delta f=\pm 8\%$.

\begin{figure}[htbp]
\centering 
\includegraphics[scale=0.36]{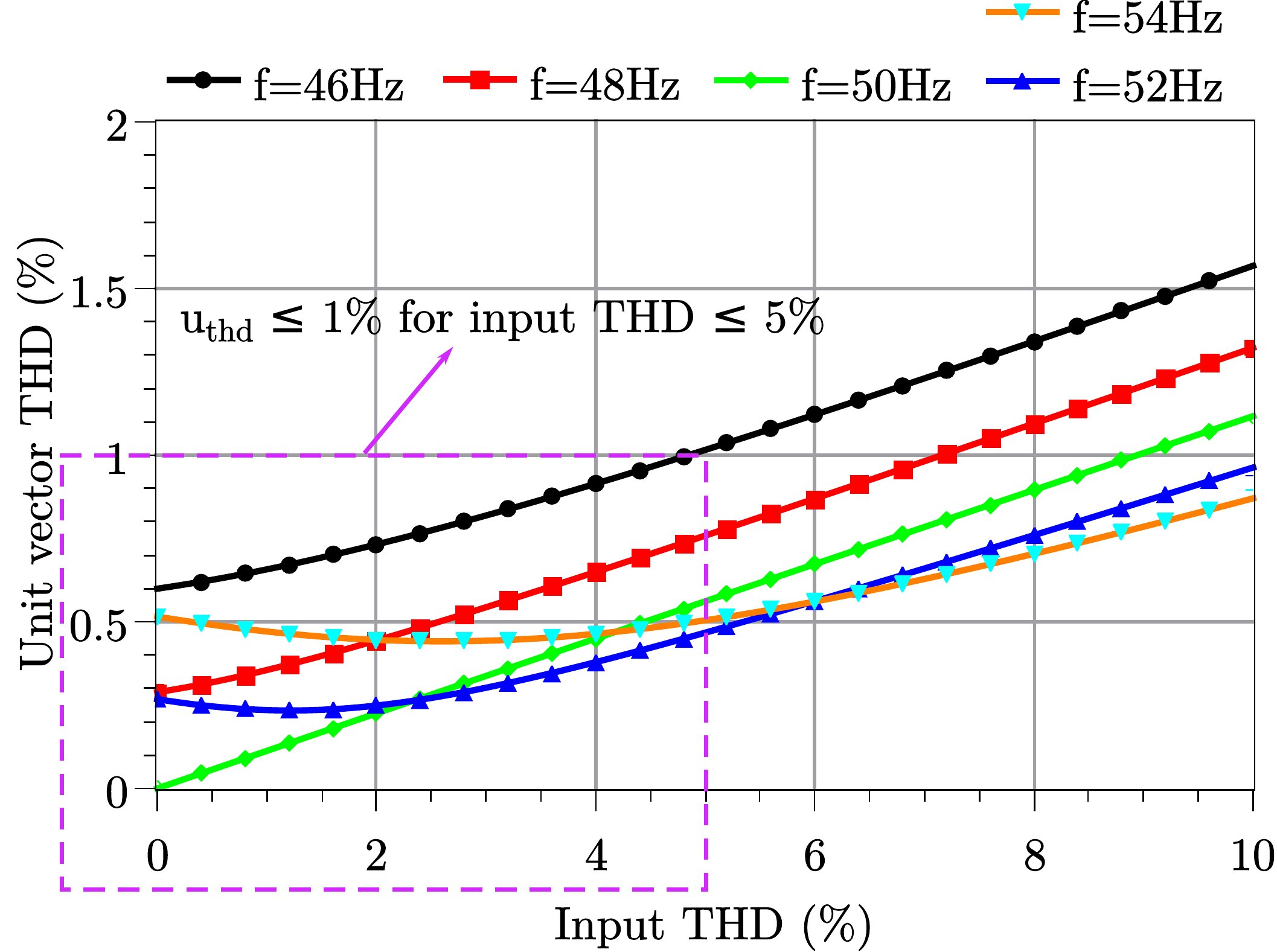} 
\caption{Variation of unit vector THD with input THD for upto $\pm8\%$ frequency deviation in the input for the proposed HC-MTSD design.}
\label{uthd_vs_ipthd2} 
\end{figure}

The variation of unit vector THD versus input THD and  input frequency deviations is shown in Fig.~\ref{uthd_vs_ipthd2} for the HC-MTSD design. 
Fig.~\ref{uthd_vs_ipthd2} can be compared with Fig.~\ref{uthd_vs_ipthd1}. It can be  seen that with the HC-MTSD design, the unit vector THD stays within $1\%$ for the worst case condition of input THD of upto $5\%$ with frequency deviation of upto $\pm 8\%$.  

\section{Design Summary}
In the proposed design methods, the parameter $k$ and the bandwidth of the embedded SRF-PLL are selected. These parameters are general and not system specific. This is because in the HGI, the transfer functions in (\ref{gatf_h}) and (\ref{gbtf_h}) are fixed. The sampling frequency and type of digital implementation determine the final difference equations. However, the value of $k$ will remain as per the proposed design in any system. Similarly, the SRF-PLL bandwidth is a general parameter. The PI controller parameters, which are calculated using the bandwidth value using (\ref{kpeqn}) -- (\ref{kieqn}), are system specific and their value will depend on discretization methods and sampling frequency. However, the bandwidth value will remain as per the proposed designs. Hence, the overall design summary  applies to any general single-phase system. It  is given as follows.
\begin{enumerate}
\item \emph{Minimum $t_{sd}$ design (MTSD design):} This design selects the HGI-PLL design parameters considering worst case input frequency deviation and limit on unit vector THD. For these conditions, this design obtains the fastest response. The design parameters determined for a frequency deviation of $\pm 8\%$ and unit vector THD limit of $1\%$ are:
\begin{align}
k=1.56,~f_{bw}=55Hz,~
t_{sd}=27.6ms \label{mtsd_summary}
\end{align} 
\item \emph{Harmonic constrained Minimum $t_{sd}$ design (HC-MTSD design):}  This design is an extension of the MTSD design. It selects the HGI-PLL design parameters considering worst case input frequency deviation, worst case input THD and a constraint on the unit vector THD. For these conditions, this design obtains the fastest response. The design parameters determined for a frequency deviation of $\pm 8\%$, input voltage THD of $5\%$ with the constraint of (\ref{harmonic_ratio}) and unit vector THD limit of $1\%$ are:
\begin{align}
k=1.56,~f_{bw}=29Hz,~
t_{sd}=37.9ms \label{hc_mtsd_summary}
\end{align} 
\end{enumerate}

The performance of the HGI-PLL is compared with popular SOGI based single-phase PLLs. The comparison is given in Table~\ref{tab_comparison}.
\newcommand\pb[1]{%
\begin{tabular}{@{}c@{}}#1\end{tabular}}
\begin{table*}[htbp]
\begin{center}
\caption{Comparison of HGI-PLL with popular SOGI based single-phase PLLs.}
\label{tab_comparison}
\begin{tabular}{|c|c|c|c|c|}
\hline
PLL Type & \pb{DC cancelling \\capability} & Design Parameters & Design Method & \pb{Resource Utilization$^\ast$  \\ (Multiplications - M and additions - A)} \\ \hline
Basic fixed SOGI-PLL~\cite{abhi_iecon} & No & 2 & Systematic & 3M, 4A \\ \hline
Basic adaptive SOGI-PLL~\cite{ciobotaru1,adaptive_sogi_blaabjerg} & No & 2 & Heuristic & 5M, 4A \\ \hline
Modified adaptive SOGI-PLL~\cite{ciobotaru2} & Yes & 3 & Heuristic & 7M, 8A \\ \hline
\pb{HGI-PLL (Present work)} & Yes & 2 & Systematic & 4M, 6A \\ \hline 
\end{tabular}\\[7pt]
\end{center}
{$^\ast$The  resource utilization in the embedded SRF-PLL  is 7M, 6A for all the SOGI based PLLs. 
}
\end{table*}
The resource utilization mentioned in Table~\ref{tab_comparison} is computed using forward Euler implementation. Trapezoidal or other discretization methods can also be used~\cite{ciobotaru1}. However, the  Euler method gives the least resources which can be important when the implementation is done using a low-end digital controller.
It can be seen from Table~\ref{tab_comparison} that HGI-PLL uses considerably lesser resource compared to the  other dc cancelling SOGI based PLL. It also has the systematic design methods for the selection of the PLL parameters.

\section{Experimental Validation}
The experimental results in this section validate the steady-state and transient performance of HGI-PLL for the following cases.
\begin{enumerate}
\item Validating the offset rejecting performance
\item Validating the transient response
\item Validating the unit vector THD  when the input contains a THD of $5\%$ along with a $-8\%$ frequency deviation.
\end{enumerate}
The experimental implementation of the HGI-PLL is done on Altera Cyclone EP1C12Q240C8 FPGA controller board using fixed point 16 bit arithmetic and a sampling rate of $50\mu s$.

\begin{figure}[htbp]
\centering 
\begin{tabular}{c}
\subfloat[]{\includegraphics[scale=0.75]{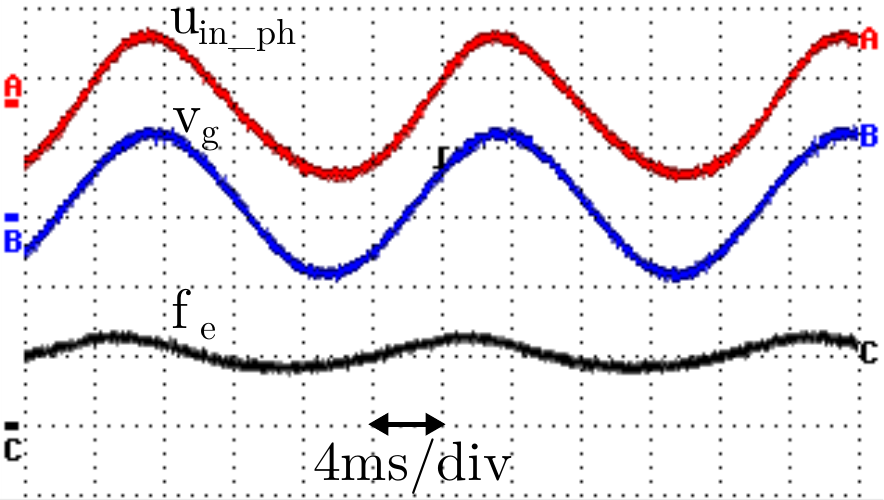}} \\
\subfloat[]{\includegraphics[scale=0.75]{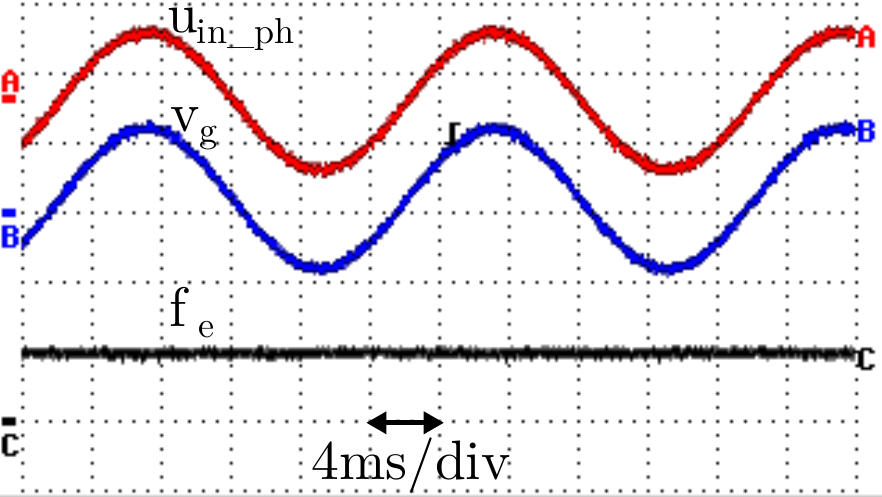}}
\end{tabular}
 \caption{Effect of a $10\%$ dc offset on (a) basic SOGI-PLL and (b) HGI-PLL. Ch. A = In-phase unit vector $u_{in\_ph}$ (1pu/div), Ch. B = Input voltage $v_g$ (5V/div), Ch. C = Estimated frequency $f_e$(50Hz/div). }
\label{dc_effect_sogipll_pllh} 
\end{figure}

Fig.~\ref{dc_effect_sogipll_pllh}(a) shows the effect of presence of $10\%$ dc offset in the input voltage of basic SOGI-PLL~\cite{abhi_iecon,ciobotaru1}. The large amount dc offset is considered to clearly show the presence of the dc offset in the input voltage.
As it can be observed from Fig.~\ref{dc_effect_sogipll_pllh}(a), the estimated frequency $f_e$ contains a ripple error at fundamental frequency. This results in the presence of dc offsets and even harmonics in the unit vector. 

The performance of HGI-PLL for the same $10\%$ input dc offset conditions is shown in Fig.~\ref{dc_effect_sogipll_pllh}(b).
The estimated frequency is a purely dc quantity for HGI-PLL indicating that the input dc has been rejected by the modified SOGI structure in  HGI-PLL.

\begin{figure}[htbp]
\centering 
\begin{tabular}{c}
\subfloat[]{\includegraphics[scale=0.75]{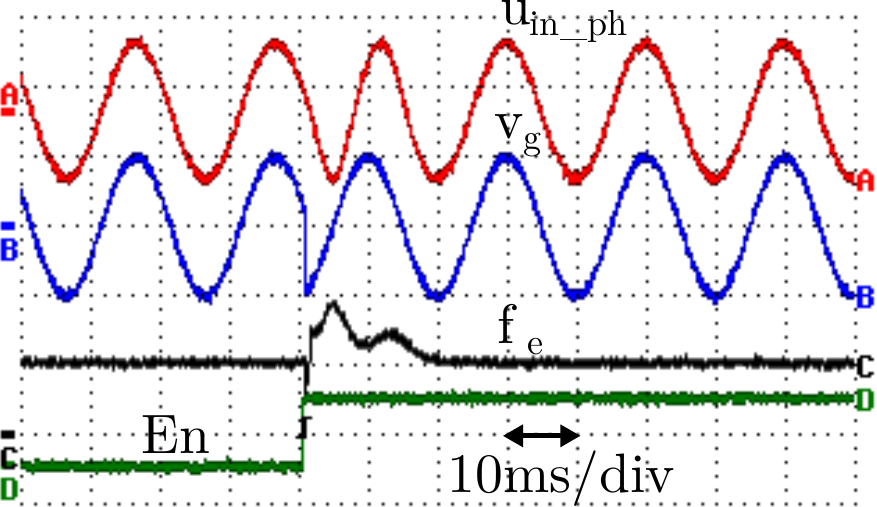}} \\
\subfloat[]{\includegraphics[scale=0.75]{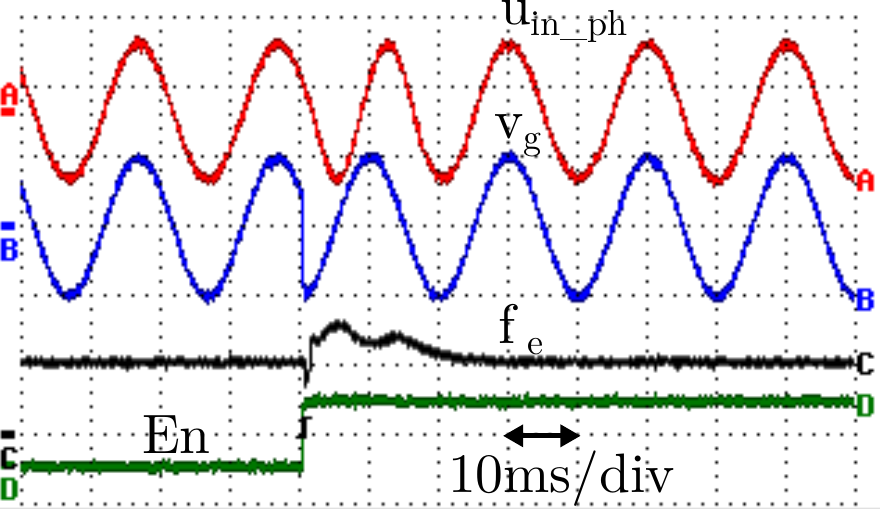}}
\end{tabular}
\caption{Transient response of HGI-PLL to step-phase-change in the input (a) MTSD design (b) HC-MTSD design. Ch. A = In-phase unit vector $u_{in\_ph}$ (1pu/div), Ch. B = Input voltage $v_g$ (5V/div), Ch. C = Estimated frequency $f_e$(50Hz/div), Ch. D = Enable (En) signal. }
\label{phase_change_pllh} 
\end{figure}

The transient response of HGI-PLL is verified by introducing a step-phase-change in the input voltage. Fig.~\ref{phase_change_pllh}(a) shows the result for MTSD design of HGI-PLL. Fig.~\ref{phase_change_pllh}(b) shows the result for HC-MTSD design of HGI-PLL. 
As it can be observed, the HC-MTSD design has slightly slower response. The settling time for MTSD design is observed to be $20ms$ whereas for the HC-MTSD design it is observed to be about $30ms$. As expected, these values are lower than the respective worst case additive-settling-time ($t_{sd}$) which were  $27.6ms$ and $37.9ms$ respectively.

\begin{figure}[htbp]
\centering 
\begin{tabular}{c}
\subfloat[]{\includegraphics[scale=0.75]{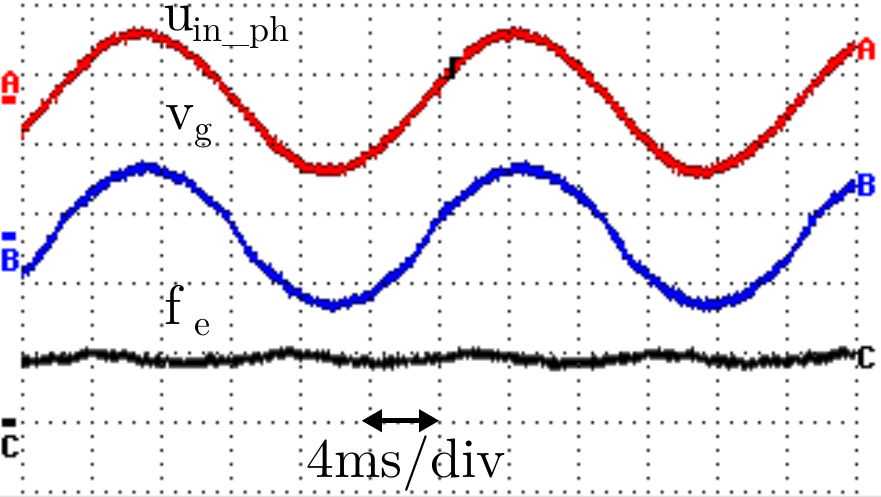}} \\
\subfloat[]{\includegraphics[scale=0.75]{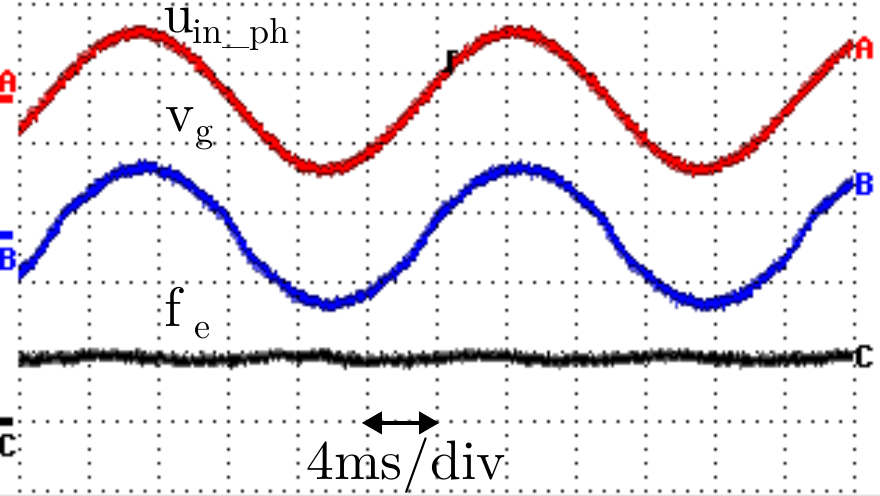}}
\end{tabular}
\caption{Effect of $-8\%$ frequency deviation in the input voltage with $5\%$ THD on HGI-PLL for (a) MTSD design and (b) HC-MTSD design.  Ch. A = In-phase unit vector $u_{in\_ph}$ (1pu/div), Ch. B = Input voltage $v_g$ (5V/div), Ch. C = Estimated frequency $f_e$ (50Hz/div). } 
\label{freq_dev_thd_perf_pllh}
\end{figure}

The effect of harmonics and frequency deviation in the input voltage is verified next. The input voltage has a fundamental frequency of $46Hz$ which corresponds to a $-8\%$ frequency deviation. The input voltage also has a $5\%$ THD. The performance of HGI-PLL with MTSD design for this input voltage is shown in Fig.~\ref{freq_dev_thd_perf_pllh}(a). The response of HGI-PLL for the same input condition with HC-MTSD design is shown in Fig.~\ref{freq_dev_thd_perf_pllh}(b). 

The time domain waveforms of the in-phase unit vector do not show significant difference in distortion. However, spectrum of the unit vectors was obtained from the experimental waveforms. The unit vector THD corresponding to the Fig.~\ref{freq_dev_thd_perf_pllh}(a) and (b) is compared with the unit vector THD from analytical results and simulation result in Table~\ref{compare_thd_pllh}(a) and (b). It can be observed that for the case of input fundamental frequency of 46Hz, the unit vector THD is reduced to less than $1\%$ when the HC-MTSD design is used. This table shows the detailed comparison between analytical, simulation and experimental results when the input THD is $5\%$ and frequency varies from 46Hz to 54Hz with a $\pm 8\%$ deviation from the nominal 50Hz case. The comparison is done for both MTSD and HC-MTSD designs.  It can be observed that the values of THD  are in agreement. The deviations observed in the experimental result are attributed  to the accuracy limitation of the oscilloscope in measuring the individual harmonics of low amplitude and the quantization errors.


\begin{table}[htbp]
\centering
\caption{Comparison of analytical, simulation and experimentally observed unit vector THD of HGI-PLL for (a) MTSD design and (b) HC-MTSD design. Input THD is 5\% with fundamental frequency variation of upto $\pm 8\%$.}
\begin{tabular}{c}
\subfloat[MTSD design]{
\begin{tabular}{|c|c|c|c|}
\cline{1-4}
 Fundamental & \multicolumn{3}{c|}{Unit vector THD (\%)}  \\ \cline{2-4}
{Frequency (Hz)} & Analytical & Simulation & Experimental \\ \hline
46 & 1.7 & 1.6 & 1.5 \\
48 & 1.3 & 1.3 & 1.3 \\
50 & 1.0 & 1.0 & 1.0 \\
52 & 0.8 & 0.8 & 0.8 \\
54 & 0.9 & 0.7 & 0.7 \\ \hline
\end{tabular}} \\
\subfloat[HC-MTSD design]{
\begin{tabular}{|c|c|c|c|}
\cline{1-4}
 Fundamental & \multicolumn{3}{c|}{Unit vector THD (\%)}  \\ \cline{2-4}
{Frequency (Hz)} & Analytical & Simulation & Experimental \\ \hline
46 & 1.0 & 0.9 & 0.9 \\
48 & 0.8 & 0.7 & 0.8 \\
50 & 0.6 & 0.6 & 0.5 \\
52 & 0.5 & 0.4 & 0.4 \\
54 & 0.5 & 0.4 & 0.4 \\ \hline
\end{tabular}}
\end{tabular}
\label{compare_thd_pllh}
\end{table}

\section{Conclusion}
In this paper, systematic designs are proposed for  high-pass generalized integrator based PLL (HGI-PLL) for single-phase grid-connected power converter application. This is a modified fixed-parameter SOGI-PLL with  input dc offset rejection capability. This property of dc rejection is important as the basic SOGI-PLLs do not have this capability and can result in dc injection to the grid when input contains dc offsets. The performance of HGI-PLL is affected by the non-ideal input conditions of  frequency deviations  as well as harmonic distortions. These non-idealities result in the harmonic distortion of the PLL unit vectors. 
This is undesirable as the unit vectors are used for the reference generation in the closed-loop control of the grid-connected power converter.

Systematic design methods are proposed in this paper for the HGI-PLL. Firstly, the HGI-PLL parameters are selected considering a worst case input frequency deviation. 
For a constraint on the unit vector THD of $1\%$, this design achieves fastest response of the HGI-PLL. This method  can exceed the limit on unit vector THD when the input contains considerable harmonic distortion. Hence, to mitigate this problem, this design is evolved to include the non-ideality of the input voltage harmonic distortion. This  is an extension of the first design and it selects the HGI-PLL parameters considering worst case frequency deviations as well as the THD in the input voltage. The design parameters are selected such that for the given worst case conditions, the HGI-PLL has the fastest response without exceeding the unit vector THD limit of $1\%$.

The proposed designs have been validated experimentally and are found to agree with the analysis. The HGI-PLL uses considerably lesser resources while being able to provide  good steady-state and transient performance. The proposed design method can be extended to arbitrary single-phase systems. HGI-PLL with the proposed designs is a suitable PLL scheme when low-end digital controllers  are used in the control of  grid-connected power converter systems, as it has low implementation complexity.

\begin{appendices}
\section{Quantifying the Effect of Frequency Deviation on Unit Vector Harmonic Distortion}
When there is a frequency deviation in the input, the amplitudes of $v_\alpha$ and $v_\beta$ become unequal as can be seen from the Bode plot in Fig.~\ref{hgipll_bode}. Let these voltages be defined as follows.
\begin{eqnarray}
v_\alpha = V_1\sin(\omega t + \phi_1)
\label{valph}  \\
v_\beta = V_2\sin(\omega t + \phi_2) \label{vbeta}
\end{eqnarray}
\end{appendices}
From the Bode plot in Fig.~\ref{hgipll_bode}, it can be deduced that $V_1\neq V_2$ in (\ref{valph}) and (\ref{vbeta}). Similarly, $\phi_1\neq \phi_2$ and $\phi_1-\phi_2=\pi /2$. These parameters are known from the transfer functions of the HGI for any given input frequency.

Due to the unequal amplitudes, the two-phase equivalent voltages $v_\alpha$ and $v_\beta$ contain an unbalance and hence a negative sequence component. This causes the well known problem of double fundamental frequency ripple in $v_d$, $v_q$ and estimated frequency $\omega _e$. As the estimated phase $\theta _e$ is integral of $\omega _e$, it will have the following form:
\begin{equation}
\theta _e = \omega _e t + f \label{error_eq}
\end{equation}
In (\ref{error_eq}), $f$ is a second harmonic ripple error. This is defined in (\ref{fdef}) below.  The aim is to determine the parameters $a$ and $\phi$ in the equation shown below, 
\begin{equation}
f=a\sin(2\omega t + \phi)
\label{fdef}
\end{equation} 
In (\ref{fdef}), $\omega$ is the fundamental frequency whose nominal value is $\omega = \omega _0 = 2\pi50~rad/s$. It is assumed that the fundamental frequency can vary upto $\pm 8\%$ in this paper. For given $f$, it is shown in~\cite{abhi1} that the unit vector will have a third harmonic amplitude equal to
\begin{equation}
u_3=\frac{a}{2} \label{third_harmonic}
\end{equation}

The detailed derivation steps to determine $a$ and $\phi$ are provided in~\cite{abhi1}. Only the final expressions are reproduced here:
\begin{align}
a =& \frac{m[(V_1/2)\cos(\phi_1+x)+(V_2/2)\sin(\phi_2+x)]}{\cos(\phi)-m\cos(\phi + x)[-\cos(\phi_1)V_1/2+\sin(\phi_2)V_2/2]} \notag \\
\phi =& \arctan\left[\frac{\alpha+\beta \nu}{\alpha \nu - \beta}\right] - x  \label{aphi} 
\end{align}

Where
\begin{eqnarray}
\alpha&=&\cos(x)+[(V_1/2)\cos(\phi_1)-(V_2/2)\sin(\phi_2)]m \notag\\
\beta&=&\sin(x) \notag\\
\nu &=& \frac{(V_1/2)\cos(\phi_1+x)+(V_2/2)\sin(\phi_2+x)}{(V_1/2)\sin(\phi_1+x)-(V_2/2)\cos(\phi_2+x)} \label{albetnu}
\end{eqnarray}

In (\ref{aphi}) and (\ref{albetnu}), $m$ and $x$ are the the overall magnitude gain and phase shift at second harmonic frequency given by the summer, PI controller and integrator in the embedded SRF-PLL in Fig.~\ref{hgipll_str}. They are expressed as follows:
\begin{align}
m&=\Bigg{|}-\left(k_p+\frac{k_i}{s}\right)\frac{1}{s} \Bigg{|}_{s=j2\omega } \notag \\
x&=\angle{\left(-\left(k_p+\frac{k_i}{s}\right)\frac{1}{s}\right)}\Bigg{|}_{s=j2\omega } \label{mx_2}
\end{align}

Thus, for a given frequency deviation, the output of the HGI $v_\alpha$ and $v_\beta$ can be determined. Then the above equations can be used to determine $a$ which equals to twice the amplitude of the third harmonic in the unit vector~\cite{abhi1}.

\section{Quantifying the Effect of Grid Voltage Harmonics on Unit Vector Harmonic Distortion}
Let the sensed grid voltage contain a harmonic of the order $h$. Depending on the transfer functions of the HGI, the voltages $v_\alpha$ and $v_\beta$ will also contain harmonic of the order $h$ whose magnitude and phase can be calculated. Let these harmonic voltages be defined in phasor form as follows:
\begin{align}
\mathbf{v_{h\alpha}}&=V_{h\alpha}\angle{\phi _h} \notag \\
\mathbf{v_{h\beta}}&=V_{h\beta}\angle{\psi _h} \label{harmonic_phasors}
\end{align}
The phasors in (\ref{harmonic_phasors}) rotate at the harmonic frequency that is $h$ times the fundamental.
These harmonic voltages can be split into positive and negative sequence voltages in two-phase system as follows:
\begin{align}
\mathbf{v_{h\alpha p}} &= \frac{\mathbf{v_{h\alpha}} + j \mathbf{v_{h\beta}}}{2} \notag \\
\mathbf{v_{h\alpha n}} &= \frac{\mathbf{v_{h\alpha}} - j \mathbf{v_{h\beta}}}{2} \label{harmonic_seq_alpha}
\end{align}
In (\ref{harmonic_seq_alpha}), the voltages $\mathbf{v_{h\alpha p}}$ and $\mathbf{v_{h\alpha n}}$ are the $\alpha$ axis positive and negative sequence voltages. The corresponding $\beta$ axis voltages are given by,
\begin{align}
\mathbf{v_{h\beta p}} &= -j\mathbf{v_{h\alpha p}}  \notag \\
\mathbf{v_{h\beta n}} &=  j\mathbf{v_{h\alpha n}}   \label{harmonic_seq_beta}
\end{align}
The expressions in (\ref{harmonic_seq_beta}) are obtained based on the fact that $\beta$ axis voltage lags $\alpha$ axis voltage by $90^\circ$ in positive sequence while it leads by $90^\circ$ in negative sequence. By adding up positive and negative sequence voltages, the original voltages in (\ref{harmonic_phasors}) can be obtained.

In~\cite{abhi1}, unit vector harmonic distortion is determined analytically when the input contains a harmonic of any given sequence. The harmonic with order $h$ occurring as  a positive sequence gives rise to $(h-2)$ and $h$ order harmonics in the unit vectors. Similarly  the harmonic with order $h$ occurring as  a negative sequence gives rise to $h$ and $(h+2)$ order harmonics in the unit vectors.

The expressions in~\cite{abhi1} are reproduced here for computing the distortion due to positive sequence harmonic. Let these voltages be:
\begin{align}
v_{h\alpha p} &= V_h\sin(h\omega t+\gamma)  \notag\\
v_{h\beta p} &= -V_h\cos(h\omega t+\gamma) \label{harmonic_timedom}
\end{align}
The expressions in (\ref{harmonic_timedom}) are the general time domain expressions of the corresponding phasors specified in (\ref{harmonic_seq_alpha}) and (\ref{harmonic_seq_beta}).
The sensed grid voltage has a positive sequence fundamental voltage given in the following general form:
\begin{align}
v_{1+(\alpha)}=&V_{1+}\sin(\omega t+\delta) \notag \\
v_{1+(\beta)}=&-V_{1+}\cos(\omega t+\delta)\label{v1+}
\end{align} 
For the input conditions as in (\ref{v1+}) and (\ref{harmonic_timedom}), the analytical expressions are derived in~\cite{abhi1} to compute the $(h-2)$ and $h$ order harmonics in the unit vector. 
The sine unit vector will have the following  form for these harmonics,
\begin{align}
{u_{h-2}}=&a_h\sin((h-2)\omega t + \phi _h) \notag \\
{u_{h}}=&a_h\sin(h\omega t + \phi _h) \label{unit_vec_harmonic}
\end{align}

The final expressions for the amplitude and phase in (\ref{unit_vec_harmonic}) are as follows.
\begin{eqnarray}
\phi_h&=&\arctan\left[\frac{\alpha _h-\beta _h \cot(x_{h-1}+\gamma)}{\beta _h+\alpha _h \cot(x_{h-1}+\gamma)}\right] \notag\\
a_h&=&\frac{0.5V_h m_{h-1}\cos(x_{h-1}+\gamma)}{\cos(\phi_h)+m_{h-1}V_{1+}\cos(\delta)\cos(\phi_h+x_{h-1})}  \label{aphikk}
\end{eqnarray}
Where
\begin{eqnarray}
\alpha _h&=&1+m_{h-1}V_{1+}\cos(\delta)\cos(x_{h-1}) \notag\\
\beta _h&=&m_{h-1}V_{1+}\cos(\delta)\sin(x_{h-1}) \label{alphbetkk}
\end{eqnarray}

In (\ref{aphikk}) and (\ref{alphbetkk}), $m_{h-1}$ and $x_{h-1}$ are the the overall magnitude gain and phase shift at second harmonic frequency given by the summer, PI controller and integrator in the embedded SRF-PLL in Fig.~\ref{hgipll_str}. They are expressed as follows:
\begin{align}
m_{h-1}&=\Bigg{|}-\left(k_p+\frac{k_i}{s}\right)\frac{1}{s} \Bigg{|}_{s=j(h-1)\omega } \notag \\
x_{h-1}&=\angle{\left(-\left(k_p+\frac{k_i}{s}\right)\frac{1}{s}\right)}\Bigg{|}_{s=j(h-1)\omega } \label{mx_h}
\end{align}
For the   negative sequence harmonic, the same expressions (\ref{aphikk})--(\ref{mx_h}) can be used by replacing $h-1$ with $h+1$. 
The overall THD is determined by performing a phasor sum of the harmonics in the unit vector appearing due to all the positive sequence and negative sequence  harmonics in the input voltage to the embedded SRF-PLL. 

\section{Expressions for the PI Controller Parameters}
For any given design bandwidth of the embedded SRF-PLL ($\omega _{bw}$), the PI controller parameters $k_p$ and $k_i$ can be determined uniquely. The corresponding equations are  as follows~\cite{blasko,abhi_iet},
\begin{align}
k_p&=\frac{\omega_{bw}}{V_m} \label{kpeqn} \\
k_i&=k_pT_s\omega_{bw}^2 \label{kieqn}
\end{align}
In (\ref{kieqn}),  parameter $T_s$ is the sampling time used in the digital implementation of the SRF-PLL. In (\ref{kpeqn}),   $V_m$ is the nominal sensed grid voltage peak. 

\bibliographystyle{ieeetr}
\bibliography{references}

\begin{thebibliography}{10}

\bibitem{pll_soc}
K.~Nagaraj, A.~S. Kamath, K.~Subburaj, B.~Chattopadhyay, G.~Nayak, S.~S. Evani,
  N.~P. Nayak, I.~Prathapan, F.~Zhang, and B.~Haroun, ``Architectures and
  circuit techniques for multi-purpose digital phase lock loops,'' {\em IEEE
  Transactions on Circuits and Systems I: Regular Papers}, vol.~60,
  pp.~517--528, March 2013.

\bibitem{pll_dg1}
H.~Tao, J.~Duarte, and M.~Hendrix, ``Control of grid-interactive inverters as
  used in small distributed generators,'' in {\em 42nd IAS Annual Meeting,
  Industry Applications Conference}, pp.~1574--1581, Sept 2007.

\bibitem{pll_dg2}
Z.~Wang, S.~Fan, Y.~Zheng, and M.~Cheng, ``Control of a six-switch inverter
  based single-phase grid-connected pv generation system with inverse park
  transform pll,'' in {\em IEEE International Symposium on Industrial
  Electronics (ISIE)}, pp.~258--263, May 2012.

\bibitem{pv_dg_monitor}
R.~Teodorescu and F.~Blaabjerg, ``Flexible control of small wind turbines with
  grid failure detection operating in stand-alone and grid-connected mode,''
  {\em IEEE Transactions on Power Electronics}, vol.~19, pp.~1323--1332, Sept
  2004.

\bibitem{statcom2}
R.~Menzies and G.~Mazur, ``Advances in the determination of control parameters
  for static compensators,'' {\em IEEE Transactions on Power Delivery}, vol.~4,
  pp.~2012--2017, Oct 1989.

\bibitem{statcom3}
A.~Norouzi and A.~Sharaf, ``Two control schemes to enhance the dynamic
  performance of the {STATCOM} and {SSSC},'' {\em IEEE Transactions on Power
  Delivery}, vol.~20, pp.~435--442, Jan 2005.

\bibitem{bsingh}
B.~Singh and S.~Arya, ``Implementation of single-phase enhanced phase-locked
  loop-based control algorithm for three-phase dstatcom,'' {\em IEEE
  Transactions on Power Delivery}, vol.~28, pp.~1516--1524, July 2013.

\bibitem{abhi1}
A.~Kulkarni and V.~John, ``Analysis of bandwidth-unit-vector-distortion
  tradeoff in pll during abnormal grid conditions,'' {\em IEEE Transactions on
  Industrial Electronics}, vol.~60, pp.~5820--5829, Dec 2013.

\bibitem{pll_monitor1}
A.~Luna, J.~Rocabert, J.~Candela, J.~Hermoso, R.~Teodorescu, F.~Blaabjerg, and
  P.~Rodriguez, ``Grid voltage synchronization for distributed generation
  systems under grid fault conditions,'' {\em IEEE Transactions on Industry
  Applications}, vol.~51, pp.~3414--3425, July 2015.

\bibitem{karimi}
M.~Karimi-Ghartemani, ``A novel three-phase magnitude-phase-locked loop
  system,'' {\em IEEE Transactions on Circuits and Systems I: Regular Papers},
  vol.~53, pp.~1792--1802, Aug 2006.

\bibitem{pll_monitor3}
C.~Fitzer, M.~Barnes, and P.~Green, ``Voltage sag detection technique for a
  dynamic voltage restorer,'' {\em IEEE Transactions on Industry Applications},
  vol.~40, pp.~203--212, Jan 2004.

\bibitem{blasko}
V.~Kaura and V.~Blasko, ``Operation of a phase locked loop system under
  distorted utility conditions,'' {\em IEEE Transactions on Industry
  Applications}, vol.~33, pp.~58--63, Jan 1997.

\bibitem{srfpll_based1}
Y.~F. Wang and Y.~W. Li, ``Grid synchronization pll based on cascaded delayed
  signal cancellation,'' {\em IEEE Transactions on Power Electronics}, vol.~26,
  pp.~1987--1997, July 2011.

\bibitem{srfpll_based3}
M.~Meral, ``Improved phase-locked loop for robust and fast tracking of three
  phases under unbalanced electric grid conditions,'' {\em IET Generation,
  Transmission Distribution}, vol.~6, pp.~152--160, February 2012.

\bibitem{ghoshal2015}
A.~Ghoshal and V.~John, ``Performance evaluation of three phase {SRF-PLL} and
  {MAF-SRF-PLL},'' {\em Turkish Journal of Electrical Engineering and Computer
  Sciences}, vol.~23, pp.~1781--1804, 2015.

\bibitem{ciobotaru1}
M.~Ciobotaru, R.~Teodorescu, and F.~Blaabjerg, ``A new single-phase pll
  structure based on second order generalized integrator,'' in {\em 37th IEEE
  Power Electronics Specialists Conference (PESC)}, pp.~1--6, June 2006.

\bibitem{ciobotaru2}
M.~Ciobotaru, R.~Teodorescu, and V.~Agelidis, ``Offset rejection for pll based
  synchronization in grid-connected converters,'' in {\em Twenty-Third Annual
  IEEE Applied Power Electronics Conference and Exposition (APEC)},
  pp.~1611--1617, Feb 2008.

\bibitem{abhi_iecon}
A.~Kulkarni and V.~John, ``A novel design method for sogi-pll for minimum
  settling time and low unit vector distortion,'' in {\em 39th Annual
  Conference of the IEEE Industrial Electronics Society-IECON 2013},
  pp.~274--279, Nov 2013.

\bibitem{adaptive_sogi_blaabjerg}
Y.~Yang and F.~Blaabjerg, ``Synchronization in single-phase grid-connected
  photovoltaic systems under grid faults,'' in {\em 3rd IEEE International
  Symposium on Power Electronics for Distributed Generation Systems (PEDG)},
  pp.~476--482, June 2012.

\bibitem{adaptive_sogi1}
P.~Rodriguez, R.~Teodorescu, I.~Candela, A.~Timbus, M.~Liserre, and
  F.~Blaabjerg, ``New positive-sequence voltage detector for grid
  synchronization of power converters under faulty grid conditions,'' in {\em
  37th IEEE Power Electronics Specialists Conference (PESC)}, pp.~1--7, June
  2006.

\bibitem{abhi_iet}
A.~Kulkarni and V.~John, ``Design of synchronous reference frame phase-locked
  loop with the presence of dc offsets in the input voltage,'' {\em IET Power
  Electronics}, vol.~8, no.~12, pp.~2435--2443, 2015.

\bibitem{dc_offset_effect_pow_del}
G.~Buticchi, E.~Lorenzani, and G.~Franceschini, ``A dc offset current
  compensation strategy in transformerless grid-connected power converters,''
  {\em IEEE Transactions on Power Delivery}, vol.~26, pp.~2743--2751, Oct 2011.

\bibitem{ieee1547}
``{IEEE} standard for interconnecting distributed resources with electric power
  systems,'' {\em IEEE Std 1547-2003}, 2003.

\bibitem{multiple_cascade}
J.~Matas, M.~Castilla, J.~Miret, L.~Garcia~de Vicuna, and R.~Guzman, ``An
  adaptive prefiltering method to improve the speed/accuracy tradeoff of
  voltage sequence detection methods under adverse grid conditions,'' {\em IEEE
  Transactions on Industrial Electronics}, vol.~61, pp.~2139--2151, May 2014.

\bibitem{robust_shinnaka}
S.~Shinnaka, ``A robust single-phase pll system with stable and fast
  tracking,'' {\em IEEE Transactions on Industry Applications}, vol.~44,
  pp.~624--633, March 2008.

\bibitem{pow_del_dc_off1}
M.~Reza, M.~Ciobotaru, and V.~Agelidis, ``Accurate estimation of single-phase
  grid voltage parameters under distorted conditions,'' {\em IEEE Transactions
  on Power Delivery}, vol.~29, pp.~1138--1146, June 2014.

\bibitem{current_control_ref1}
A.~Timbus, M.~Liserre, R.~Teodorescu, P.~Rodriguez, and F.~Blaabjerg,
  ``Evaluation of current controllers for distributed power generation
  systems,'' {\em IEEE Transactions on Power Electronics}, vol.~24,
  pp.~654--664, March 2009.

\bibitem{pcc_dist1}
X.~Zong, P.~Gray, and P.~Lehn, ``New metric recommended for {IEEE Std. 1547} to
  limit harmonics injected into distorted grids,'' {\em IEEE Transactions on
  Power Delivery}, vol.~PP, no.~99, pp.~1--1, 2015.

\bibitem{pcc_dist2}
``{IEEE} recommended practices and requirements for harmonic control in
  electrical power systems,'' {\em {IEEE} Std 519-1992}, pp.~1--112, April
  1993.

\bibitem{ogata}
K.~Ogata, {\em Modern Control Engineering}.
\newblock Prentice Hall, 5th~ed., 2008.

\bibitem{ieee1159}
``{IEEE} recommended practice for monitoring electric power quality,'' {\em
  {IEEE Std} 1159-2009 (Revision of {IEEE Std} 1159-1995)}, pp.~c1--81, June
  2009.

\end{thebibliography}

\end{document}